\documentclass[pdftex,twocolumn,epjc3]{svjour3}          

\RequirePackage[T1]{fontenc}

\smartqed  

\RequirePackage{graphicx}
\RequirePackage{mathptmx}      
\RequirePackage{flushend}
\RequirePackage[numbers,sort&compress]{natbib}
\RequirePackage[colorlinks,citecolor=blue,urlcolor=blue,linkcolor=blue]{hyperref}
\usepackage{float}
\usepackage{graphicx}
\usepackage{amsmath}
\usepackage{dblfloatfix}
\usepackage{multicol}
\usepackage{color}
\usepackage{xcolor}
\usepackage{amsfonts} 
\usepackage{multirow}

\journalname{Eur. Phys. J. C}

\begin{document}
	
\title{Finslerian extension of an anisotropic strange star in the domain of modified gravity}

\author{Sourav Roy Chowdhury\thanksref{addr1,e1}, Debabrata Deb\thanksref{addr2,e2}, Farook Rahaman\thanksref{addr3,e3}, Saibal Ray\thanksref{addr4,e4}} 

\thankstext{e1}{e-mail: b.itsme88@gmail.com}
\thankstext{e2}{e-mail: d.deb32@gmail.com}
\thankstext{e3}{e-mail: rahaman@associates.iucaa.in} 
\thankstext{e4}{e-mail: saibal.ray@gla.ac.in }

\institute{Department of Physics, Vidyasagar College, kolkata-700157, West Bengal, India\label{addr1} 
\and The Institute of Mathematical Sciences, CIT Campus, Taramani, Chennai 600113, Tamil Nadu, India\label{addr2} \and Department of Mathematics, Jadavpur University, Kolkata 700032, West Bengal, India\label{addr3} \and Center for Cosmology, Astrophysics and Space Science (CCASS), GLA University, Mathura 281406, Uttar Pradesh, India\label{addr4}}

\date{Received: date / Accepted: date}

\maketitle

\begin{abstract}
In this article, we apply the Finsler spacetime to develop the Einstein field equations in the extension of modified geometry. Following Finsler geometry, which is focused on the tangent bundle with a scalar function, a scalar equation should be the field equation that defines this structure. This spacetime maintains the required causality properties on the generalized Lorentzian metric manifold. The matter field is coupled with the Finsler geometry to produce the complete action. In this work, we use modified gravity to develop the Einstein field equations from the variational principle. Developed Einstein field equations are employed on the strange stellar system to improve the study. The interior of the system is made of a strange quark, maintained by the MIT Bag equation of state. In addition, the modified Tolman-Oppenheimer-Volkov (TOV) equation is formulated. In particular, the anisotropic stress attains the maximum at the surface. The mass-central density variation justifies the stability of the system.
\end{abstract}


\section{Introduction}	

General relativity (GR) is based on a spacetime manifold furnished with the metric tensor consisting of the Lorentzian signature. The Einstein equations are determined from the metric. The geodesic equation of the system figures out the motion of the particles. Different characteristics and behaviour of spacetime in four dimensions together with higher dimensions have been studied by several theoretical physicists~\cite{Harko1,Chakraborty,vacaru1,Pfeifer12,Li2014}. Most investigations are concentrated on spacetime, especially the Einstein field equations. There are diverse reasons for an investigation of gravitational theories. Plenty of them are immense on the theoretical predictions, pointed out that GR should be superseded in a more general aspect and a few tested results. In addition to astrophysical relevance, there are vast applications of spacetime to de Sitter gauge theory, induced gravity, string theory, and ADS/CFT correspondence \cite{Banados1998,Nojiri2004,Zet2006}.

To realize the formation and evolution of galaxies, figure out the dynamics and morphologies of galaxies, early-stage formation of the universe, reionisation of the universe, the rotational curves of galaxies, formation of stars, different stages of stars, the merger of the binary compact objects, several observational, simulation and experimental studies are being pursued~\cite{Baiotti,East,Peloso,Gnedin,Bouwens,Riess,Peebles,Corbelli}. The parametrized post-Newtonian (PPN) formalism is a mathematical mechanism to find the deviation of GR and the experimental results~\cite{Hohmann2014,Avilez2015,Viraj2017}. However, the PPN formalism is confined to only metric theories of gravitation. Observations and measurements of the magnitude and redshift of supernovas are also innovative studies~\cite{Kowalski2008,Amanullah2010,Suzuki2012}. The outcomes of the data analysis of supernovae reveal that, at present, the decelerating parameter ($q$) lies in the domain of $-1.0 \leq q \leq -0.5$ ~\cite{Cunha2009,Li2011}. The accelerating phase of the universe is confirmed by the negative values of the decelerating parameter. The net outcomes of these results are the contrary behaviour described by GR. 

The observation of the motion of point particles provides an adept explanation of the physical properties of spacetime. The geodesic of geometry of spacetime can be considered as the observed trajectory. The appropriateness of geometry can be verified from the matching of predicted geodesic with the observed curve. Finsler geometry is the geometry where the manifold accounted with the Finsler function. It has a 1-homogeneous function on the tangent bundle of spacetime, and the length measure for curves is associated with the Lorentzian metric. Generalized expression of the length for a curve on a manifold generalized the metric geometry~\cite{Asanov,Bao2000,Mo2000}
\[
S[\gamma] = \int d \tau \mathcal{F}(x(\tau), \dot{x}(\tau)).
\] 

Lorentzian metric $g_{ab} $ determines the function $\mathcal{F}$ as $\mathcal{F}(x,\dot{x}) = |g_{ab}(x)\dot{x}^a \dot{x}^b|^{1/2} $.  Here, $ x $ and $\dot{x}$ stand for position vector and tangent vector, respectively. According to the definition, the geometric fields along with the objects not merely depend on the points of the manifold but also depend on its tangent directions.

It is impossible to describe the dynamic physical processes without a clock. Due to the versatility of the postulate of the Finsler clock, Finsler geometry has emerged in different contexts to describe physics. The importance of Finsler geometry has been realized and explored from the viewpoint of fundamental approaches of theoretical physics: from different aspects of different gravity theories; hyperbolic polynomials defined in the generalized backgrounds~\cite{Gibbons}; to define the modified dispersion relations of Planck scale in the effective classical geometry and the corresponding breaking of local Lorentz invariance~\cite{Amelino,Gambini,Girelli}, and in linear as well as nonlinear optical media the covariant formulations
for electrodynamical system~\cite{Raetzel,Hehl}. The Finslerian extension also provides an acceptable explanation of the anomaly that GR does not fulfil for the phenomenological level, astronomical and
cosmological data, dark matter and dark energy~\cite{Chang8,Chang9,Li10}.

Modification of GR introduced by Buchdahl~\cite{Buchdahl} by introducing Ricci scalar $R$ as an arbitrary function $f(R)$ in the Einstein-Hilbert action and followed by a few more literatures in this gravity~\cite{Nojiri1,Carroll}. Later on different modifications have been done by modifying the geometric term of the action:  $f(G)$ gravity~\cite{Bamba,Rodrigues},  $f(\mathbb{T})$ gravity \cite{Bengochea,Bodmer},  $f(R, G)$ gravity~\cite{Nojiri2} and so on (where $G~\textrm{and}~\mathbb{T}$  are the Gauss-Bonnet scalar and torsion scalar, respectively). Regarding the validation of $f(R)$ gravity, it fails to uphold the Solar system tests \cite{Erickcek,Capozziello}. The gravity is also unaccounted for the stable stellar configuration~\cite{Briscese,Kobayashi}. In addition, the scalar-tensor gravity and $f(R)$ gravity are classically equivalent to each other~\cite{Fujii,Ketov}. In non-modified frame several has studied strange stellar objects \cite{np1,Sourav,Sourav2,np3,np4}.

Including the matter Lagrangian with any arbitrary form of Ricci scalar $(R)$ as well as the trace of energy momentum tensor $(\mathcal{T})$ Harko et al.~\cite{Harko1} define the $f(R, \mathcal{T})$ gravity. Theoretical divisions of cosmology along with astrophysics have successfully been studied the $f(R, \mathcal{T})$ gravity~\cite{Myrzakulov,Jamil,Shabani1,Shabani2,Moraes1,Zaregonbadi,Shabani3}. For the self-gravitating, the spherically symmetric system, the effects of the stability for the locally isotropic system has been explored by Sharif et al.~\cite{Sharif}. Noureen et al. introduce the perturbation effects in the system. Several scientists have also studied different characteristics of dynamical instability of spherically symmetric anisotropic collapsing stars~\cite{Noureen1,Noureen2,Noureen3,Zubair}. Palatini approach of $f(R, \mathcal{T})$ gravity, independent of metric, are presented in the literature~\cite{Wu,Barrientos}. The hydrostatic equation for the stellar system for isotropic together with anisotropic system reviewed by Moraes et al.~\cite{Moraes2} and Deb et al.~\cite{Deb}. Coupling of matter with curvature reveals that the energy momentum tensor is non-conserved ($\nabla_\mu T^{\mu \nu} \neq 0$) \cite{Harko1,Barrientos1}, i.e., the presence of the additional force due to the coupling. Hence, the conclusion can be drawn that gravity violates the equivalence principle of GR~\cite{Damour}. 

Interestingly, Chakraborty~\cite{Chakraborty1} considering the action (coupling of the matter and geometry) maintained the restriction to the case where the test particle moves along a geodesic. As a consequence, it is shown that the originated matter from two non-interacting fluids within the stellar system holds conservation of the effective energy-momentum tensor. From the modification of gravitation, Lagrangian introduces the supplementary force in $f(R, \mathcal{T})$ gravity which is also used to stabilize the stellar system in addition with the hydrodynamic force, anisotropic force and gravitational force~\cite{Deb}. Shabani and Farhaudi~\cite{Shabani2} provide the consequences of cosmological and Solar system in the $f(R, \mathcal{T})$ gravity, which were consistent with the observational data. The justification of the dark matter galactic effects and the gravitational lensing also support the validity of the modified theory~\cite{Zaregonbadi1}. In Finsler spacetime, modified gravity applications studied in the literatures~\cite{Vacaru3,Vacaru4}. 

In this article, we define the Finsler structure based on the following characteristics:

(i) Finsler function is the fundamental variable of the geometry that has a homogeneous scalar equation on the tangent bundle.

(ii) The geometric structure is constructed from the Finsler function and is of simplified form.

(iii) The fundamental dynamical variable is no more the metric in the Finsler geometry as it is in the semi-Riemannian geometry.

(iv)  By the variation of the action integral, the field equation is obtained.

(v) For pseudo-Riemannian geometry, the Finslerian spacetime geometry becomes similar to the dynamics determined from the Einstein field
equations.

(vi) The modified gravity maintains the system.

(vii) The interior of the stellar system is made up of the up ($u$), down ($d$) and strange ($s$) quarks, respectively and the matter distribution is maintained by the phenomenological MIT bag EOS.

From the physics point of view, the geometry is a non-metric spacetime geometry, but the main motivation of considering is that it introduces the intrinsic local anisotropy. This anisotropy contributes to the structure of astrophysical objects through the so called \textit{Finslerian parameter}. Plan of the present study is in the following order: 

The concise definition of the Finsler spacetime in Sec. \ref{sec a}. In Sec. \ref{sec b} we introduce the action principle and complete gravity equation including matter and as the stage where we have developed the Einstein field equation for modified gravity with the Finslerian background. The quadratic form of the Finsler structure is a semi-definite Riemannian structure. We have shown its consistency with the Einstein field equations. We reviewed the formation of the stellar system and hydrostatic equation with the MIT bag model equation of state (EOS) in Sec. \ref{sec 3}. Sec. \ref{dis} is devoted to the discussion of the stellar system and the generalized mass-radius limit for strange stellar configurations. In Appendix \ref{app}, we explain the constant flag in two-dimensional Finsler space.

\section{Finsler spacetime}

The basic notion in generalization of modified gravity presented in this article is based on the description of spacetime. In Sec. \ref{sec a}, we have followed up the basic notion of geometry on the Finsler spacetime~\cite{Pfeifer11}. The generation of Lorentzian metric spacetime has been introduced here. In Sec. \ref{sec b}, we have developed the Einstein field equations for the modified theory of gravity, i.e., the Lagrangian density is any arbitrary function of Ricci scalar and trace of the energy momentum tensor in the Finslerian extension. The field equation developed from the action principle, where the total action is the combination of matter and geometry.

\subsection{The Definition} \label{sec a}

The definition of Finsler spacetime has been generalized in the literatures~ \cite{Pfeifer12,Lammerzahl,Javaloyes,Hohmann} from the original definition of Beem~\cite{Beem} which are formulated and employed to describe the variety of indefinite Finsler length. 

A Finsler spacetime ($\mathbb{M}, L$) is a four dimensional smooth manifold where $L:TM \rightarrow \mathbb{R}  $ is a continuous function on the tangent bundle, known as the Finsler-Lagrange function, which satisfies the following properties:

(i) $ L $ is positive homogeneous, which has degree two, with respect to the fiber coordinates of TM.  

(ii) $ L $ is reversible in the sense $ |L(x, -y)| = |L(x,y)|$.

(iii) The Euler-Lagrange equation $\frac{d}{d \mathcal{T}} \dot{\partial _i} L - \partial_i L =0$.

For every initial condition $(x, \dot{x}) \in \mathcal{T} \cup \mathcal{N}$ there exist a unique solution, with $\mathcal{N}$ the kernel of $ L $.

(iv) $L$ is smooth and respect to the fiber coordinate, the Hessian $g^L_{ab}$ of $L$ so that $ g^L_{ab} = \frac{1}{2} \partial_a \partial_b  L$.

(v) For the preimage $L^{-1}(0, \infty) \subset TM$, there is a connected component $\mathcal{T}$, such that on  $\mathcal{T}$ the smooth $g^L$ exist with Lorentzian signature (+,~-,~-,~-).

The essence of four conical sub-bundles of $ TM \setminus\{0\} $ originated from the difficulty in defining Finsler spacetime. This characterizes the properties of the indefinite Finsler geometry as follows: 

(a) $\mathcal{N}$ is the sub-bundle where $L = 0$  and the fiber  $\mathcal{N}_x = \mathcal{N} \cap T_xM$.

(b) $\mathcal{A}$ is the sub-bundle with smooth $L$ and non-degenerate $g^L$ where the fiber is $\mathcal{A}_x = \mathcal{A} \cap T_xM$ is known as the set of admissible vectors.

(c) $\mathcal{A}_{0} = \mathcal{A} \setminus \mathcal{N}$ is the sub-bundle where $L$ is used for normalization with the fiber $\mathcal{A}_{0x} = \mathcal{A}_{0} \cap T_xM$.

(d) $\mathcal{T}$ is the conic sub bundle where $L > 0$ and the fiber $\mathcal{T}_x = \mathcal{T} \cap T_xM$. The signature of the $L$ metric is the Lorentzian signature (+,~-,~-,~-).

The extensive section is the assurance of the existence of the convex cone  $\mathcal{T}_x$ in each tangent space $T_xM$ from the definition of $\mathcal{T}$. The convexity of the $\mathcal{T}_x$ is elaborately studied in the literature \cite{Pfeifer12}. The interrelations, such as $\mathcal{A}_{0} \subset \mathcal{A}$ and $\mathcal{T} \subset \mathcal{A}_{0}$ differentiate the earlier definitions of Finsler spacetime. There is no correlation between  $\mathcal{N}$ and  $\mathcal{A}$, thus we can consider that the $L$ is not differentiable along the  direction  $L(x,\dot{x}) \neq 0$ as well as $L(x,\dot{x}) = 0$ \cite{Hohmann}.

\subsection{Basic Formalism} \label{sec b}

The Einstein field equations can be derived from the action principle.  The action is integral of the Lagrangian density over spacetime. Total action can be defined as a combination of the matter and the Einstein-Hilbert action, which couples gravity to matter as follows
\begin{eqnarray}
S =  kS_{EH} + S_{M}. \nonumber 
\end{eqnarray}

Let us now consider a Finsler space ($\mathbb{M}, \mathcal{F}$). In the Finslerian language, the Einstein-Hilbert action can be considered over the sphere bundle $\Sigma$ given by
\begin{equation}
S_{EH} = \int_{\Sigma} d^4{\hat{x}} d^3 \theta \sqrt{g} \sqrt{h} (f_{ab}y^a y^b)|_{\Sigma}.
\end{equation}

The restriction is not required of $\sqrt{g}$ and $\sqrt{h}$ to $\Sigma$. During calculation, we omit the subscript $|_{\Sigma}$ for the restriction of the functions to $\Sigma$, and all functions are meant to be evaluated there.

All quantities in the action are a function of $\mathcal{F}$ respect of g. The action, which is the integral of Lagrangian density, is varied with respect to $\mathcal{F}$. The dynamics of $\mathcal{F}$, which describes the equation of motion, are equivalent to the Einstein equations:
\begin{eqnarray}
& &\delta S_{EH}=  \int d^4{\hat{x}} d^3 \theta \sqrt{g} \sqrt{h} \Big( \frac{1}{2}f_{ab} 
g^{ab}\delta g_{ab}  +f_R \delta R_{ab} \nonumber\\& &~~~~~~~~~~~~ +\frac{1}{2}f_{ab} h^{ab}\delta h_{ab}- 2 f_R R_{ab}\frac{\delta \mathcal{F}}{\mathcal{F}} \nonumber\\& &~~~~~~~~~~~~+ 3f_{\tau}(T_{ab}-g_{ab}L_m)) \delta g_{ab} \Big)y^ay^b, \label{1}
\end{eqnarray} 
where $f_R = \partial f / \partial Ric$,  $f_\tau = \partial f / \partial \tau$ and over the sphere bundle the function $f = f_{ab}y^ay^b$~\cite{Bao2000,Bao1}. Here, $\sqrt{g}$ and $R_{ab}$ are independent of variation of $\theta$.

The variation of $h^{ab}\delta h_{ab}$ can be defined as 
\begin{equation}
h^{ab}\delta h_{ab} =  (g^{ab} - y^a y^b) \delta g_{ab} - 6 \frac{\delta \mathcal{F}}{\mathcal{F}}.\label{2}
\end{equation}

The correlation of $\delta g_{ab}$ and $\frac{\delta \mathcal{F}}{\mathcal{F}}$ can be written as $\delta g_{ab}(\hat{x}) = 2 g_{ab} \frac{\delta \mathcal{F}}{\mathcal{F}}$. On substituting Eq. (\ref{2}) and the correlations in the variational Eq. (\ref{1}) , we have
\begin{eqnarray}
& & \delta S_{EH} = \int_{\Sigma}d^4{\hat{x}} d^3 \theta \sqrt{g} \sqrt{h} \Big(2fg_{ab} - 6f_{ab}\nonumber\\
& & {\hspace{2cm}} +6f_{\tau}(T_{ab}-g_{ab}L_m) \Big)  y^ay^b\frac{\delta \mathcal{F}}{\mathcal{F}}.
\end{eqnarray}

The matter action of the Finsler space is based only on the Lagrangian density ($\textit L$) of the system, which is a scalar on the space. Therefore, we can consider that it only depends on the manifold geometry. In the Finslerian setting with Finsler function $ (\mathcal{F}) $ considered as a fundamental variable that determines spacetime, the matter action for matter fields $\psi_i$ looks like
\begin{equation}
S_{M} = \int_{\Sigma} d^4{\hat{x}} d^3 \theta \sqrt{g} \sqrt{h} L (g,\psi_i).\nonumber
\end{equation}

Due to the independency of $\textit L$ and g on $\theta$ over the fiber coordinates, on the manifold $\mathbb{M}$, we can integrate the system, which leads to the standard matter action if we divide out the volume of the three sphere.

The energy momentum tensor of the matter under consideration can be defined as the calculus of variation of the matter action with respect to the metric. In the Finsler setting, the variation with respect to the Finsler function leads to an expression that involved the Energy-Momentum tensor of $ p $-form fields on Lorentzian metric spacetime as $T^{ab}$ as well as its trace $ T = T^{ab}g_{ab} = 4 L+ 2g_{ab} \frac{\partial L}{ \partial g_{ab}} $, following Pfeifer and Wohlfarth \cite{Pfeifer12}. The variation with respect to the Finsler function is as follows
\begin{equation}
\delta S_{M} = \int_{\Sigma}d^4{\hat{x}} d^3 \theta \sqrt{g} \sqrt{h} (12T_{ab}-2Tg_{ab})y^ay^b\frac{\delta \mathcal{F}}{\mathcal{F}}.
\end{equation}

Combining the Einstein-Hilbert action with the matter leads to the total action that couples gravity to matter as follows
\[
S[F,\psi_i] = k S_{EH} + S_M.
\]

After performing the variation with respect to $\mathcal{F}$
\begin{eqnarray}
\delta S[\mathcal{F},\psi_i] 
& & {\hspace{-0.5cm}} = k \delta S_EH + \delta S_M \nonumber\\
& &  {\hspace{-0.5cm}} = \int_{\Sigma}d^4{\hat{x}} d^3 \theta \sqrt{g} \sqrt{h} \Big(k(2fg_{ab} - 6f_{ab}\nonumber\\
& & {\hspace{-2cm}} +6f_{\tau}(T_{ab}-g_{ab}L_m)) + (12T_{ab}-2Tg_{ab})\Big)  y^ay^b\frac{\delta \mathcal{F}}{\mathcal{F}}.
\end{eqnarray}

The following equation leads to determine the structure of spacetime
\begin{eqnarray}
\Big( (3f + Ric)g_{ab} - 6f_{ab} +6f_{\tau}(T_{ab}-g_{ab}L_m)\Big)y^ay^b \nonumber\\ {\hspace{4cm}}= - \frac{12T_{ab} }{k} y^ay^b. \label{5}
\end{eqnarray}

The tensors in the bracket are $y$ independent due to consideration of the space with vanishing Cartan tensor.

We consider $f = Ric + 2\eta \tau $, linear combinational form of $Ric$ and $\tau$, with a constant $\eta$ as adopted by Harko et al. \cite{Harko1}. In this study, we assume ${\textit L_m = -\mathcal{P}}$, with ${\mathcal{P} = \frac{1}{3}(p_r+2p_t)}$ and set the coupling constant $k = \frac{c^4}{4 \pi_F G}$.

The second derivative of Eq. (\ref{5}) with respect to fiber coordinates results in the Einstein equations as follows 
\begin{equation}
R_{ab}-\frac{1}{2}Rg_{ab}=\frac{8\pi_F G}{c^4}T_{ab} +\eta(\tau g_{ab}+2g_{ab} \mathcal{P}).
\end{equation}

The effective energy momentum tensor of the system can be defined as 
\begin{equation}
T^{eff}_{ab} = T_{ab} +\frac{\eta c^4}{8 \pi_F G}(\tau g_{ab}+2g_{ab} \mathcal{P}).
\end{equation}

Hereafter, we shall consider the geometrized unit, i.e., $ G=c=1 $.

Now, the covariant divergence of the stress-energy tensor is
\begin{equation}
\nabla ^a T_{ab}  = - \frac{\eta }{8\pi } \Bigl\{ g_{ab} \nabla ^a \tau + 2 \nabla^a (g_{ab} \mathcal{P})\Bigr\}.
\end{equation}

Following the above, we can write 
\begin{equation}
{T_{eff}}_{b;a} ^a= 0. \nonumber
\end{equation}

\section{Basic Equation for the stellar system} \label{sec 3}

To define the stellar structure, we assume the Finsler structure is of the form
\begin{equation}
\mathcal{F}^2= -e^{\lambda(r)} y^t y^t + e^{\nu(r)} y^r y^r +r^2 \overline{\mathcal{F}}^2(\theta, \phi, y^\theta, y^\phi ). \label{3}
\end{equation}

The metric structure coefficient can be written as
\begin{equation}
g_{\mu \nu} = \frac{\partial }{\partial y^\mu} \frac{\partial }{ \partial y^\nu} \left( \frac{1}{2} \mathcal{F}^2 \right),  \nonumber \\
\end{equation}
where $(g ^{\mu \nu} )$ = $ (g_{\mu \nu})^{-1} $ and also note that each $ g_{\mu \nu} $ is homogeneous of degree zero in $y$. 

For a non-zero vector $ y=y^{\mu}(\frac{\partial}{\partial x^\mu})\mid_{p} \in T_{p} \mathbb{M}, ~\mathcal{F} $ induces an inner product on $ T_{p} \mathbb{M} $ which is given by
\[
g_{y}(u,v)=g_{\mu \nu}(x,y)u^\mu v^\nu,
\]
where $ u=u^\mu(\frac{\partial}{\partial x^\mu})\mid_{p},~ v=v^\mu (\frac{\partial}{\partial x^\mu})\mid_{p}  \in T_{p} \mathbb{M} \setminus \lbrace 0 \rbrace$. 

Hence, the metric potential of the system can be defined as 
\[g_{\mu \nu} = diag (-e^{\lambda(r)}, e^{\nu(r)},r^2 \overline{g_{ij}}),\]
the term $\overline{g_{ij}}$ arises from ${\overline{\mathcal{F}}^2}$.

The energy momentum tensor of the anisotropic system can be considered in the following form 
\begin{equation}
T^{\mu}_{\nu}= -(\rho +p_t)u^{\mu}u_{\nu} + p_t \delta^{\mu}_{\nu} + (p_t-p_r)v^{\mu}v_{\nu}, \label{14}
\end{equation}
with $ u_{\nu} $ and $ v_{\nu} $ as the four-velocity and radial four-vector, respectively. The energy density, radial and tangential pressures of the anisotropic fluid are successively represented by $\rho$, $p_r$ and $p_t$.

The Einstein field equations for an anisotropic stellar system are in the form
\begin{align} \centering
&\frac{\nu' e^{-\nu}}{r} - \frac{ e^{-\nu}}{r^2} +\frac{\overline{Ric}}{r^2} = 8 \pi_F \Big(\rho +\frac{\eta}{24 \pi_F }(3 \rho -p_r -2p_t)\Big) \nonumber \\&{\hspace*{6cm}} = 8 \pi_F \rho^{eff}, \label{15a}\\
&\frac{\lambda' e^{-\nu}}{r} + \frac{ e^{-\nu}}{r^2} - \frac{\overline{Ric}}{r^2} = 8 \pi_F \Big(p_r -\frac{\eta}{24 \pi_F }(3 \rho -p_r -2p_t)\Big) \nonumber \\&{\hspace*{6cm}} = 8 \pi_F p_r^{eff}, \label{16a}\\
&e^{-\nu}\left[ \frac{\lambda''}{2}+\frac{\lambda'^2}{4}-\frac{\lambda' \nu'}{4} +\frac{\lambda'-\nu'}{2r} \right] \nonumber \\&{\hspace*{1cm}}= 8 \pi_F \Big(p_t  -\frac{\eta}{24 \pi_F }(3 \rho -p_r -2p_t)\Big) = 8 \pi_F p_t^{eff}, \label{17a} 
\end{align}
where $\overline{Ric}$ represents the Ricci scalar, derived from $\overline {\mathcal{F}}^2$.

To define the strange stellar system, we consider monotonically decreasing non-singular matter density within the spherically symmetric system as considered by Mak and Harko~\cite{Mak2002}, in the following form
\begin{equation}
\rho(r)=\rho_c\left[1-\left(1-\frac{\rho_0}{\rho_c}\right)\frac{r^{2}}{R^{2}}\right],\label{19}
\end{equation}
where $\rho_c$ and $\rho_0$ are the central and surface densities, respectively. 

\begin{figure}[htp!]
	\includegraphics[scale=0.4]{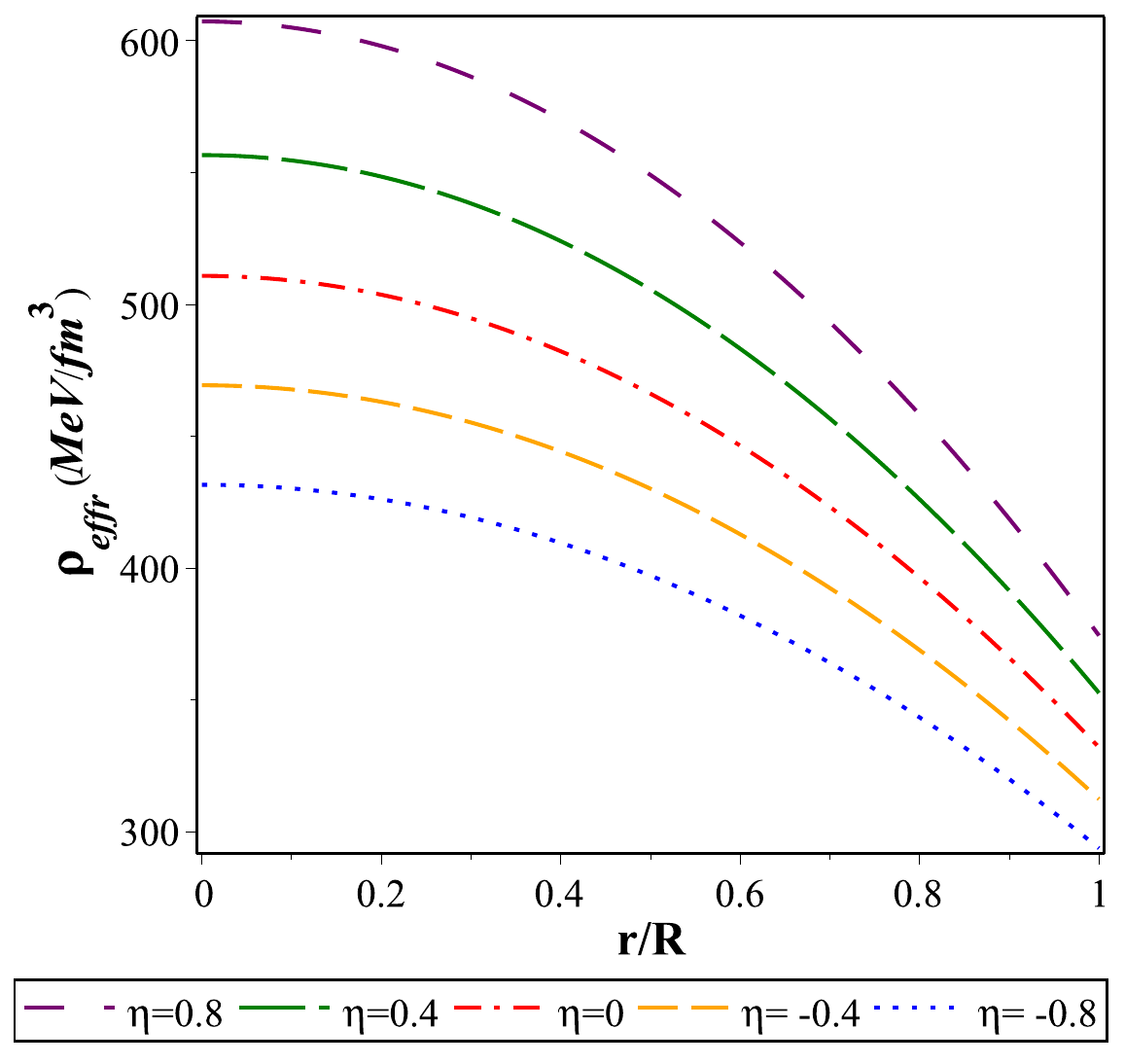} 
	\caption{Variation of the density $(\rho)$ as a function of the fractional radial coordinate $r/R$, with bag constant (B$_g$) = 83~MeV/fm$^3$ and Finsler parameter ($\overline{Ric}$) = 1.2 for the $LMC~X-4$.}\label{fden}
\end{figure}

Fig.~\ref{fden} shows the variation of density with the fractional radial function for different coupling constants.

We presumed the internal matter distribution of the strange stellar system is defined by the phenomenological MIT bag model EOS followed by Chodos et al.~\cite{Chodos1974}. The three flavoured quarks considered as the basic foundation of the bag are regarded as massless and non-interacting. Following this, the total quark pressure can be assumed as
\[
p_r = {\sum_f}{p^f} - {B_g},
\]
where $ p^f $ represents the pressure of the up ($u$), down ($d$) and strange ($s$) quarks successively, and the vacuum energy density (also known as bag constant) of the system is $B$. Here, the pressure of individual quarks is related to the energy density $ {{\rho}^f}$  of the individual quarks in the following manner $ p^f =\frac{1}{3}{{\rho}^f}$. 

The energy density of each de-confined quark is as follows
\[
{\sum_f}{{\rho}^f}=\rho+B_g.
\]

\begin{figure}[h!]
	\includegraphics[scale=0.4]{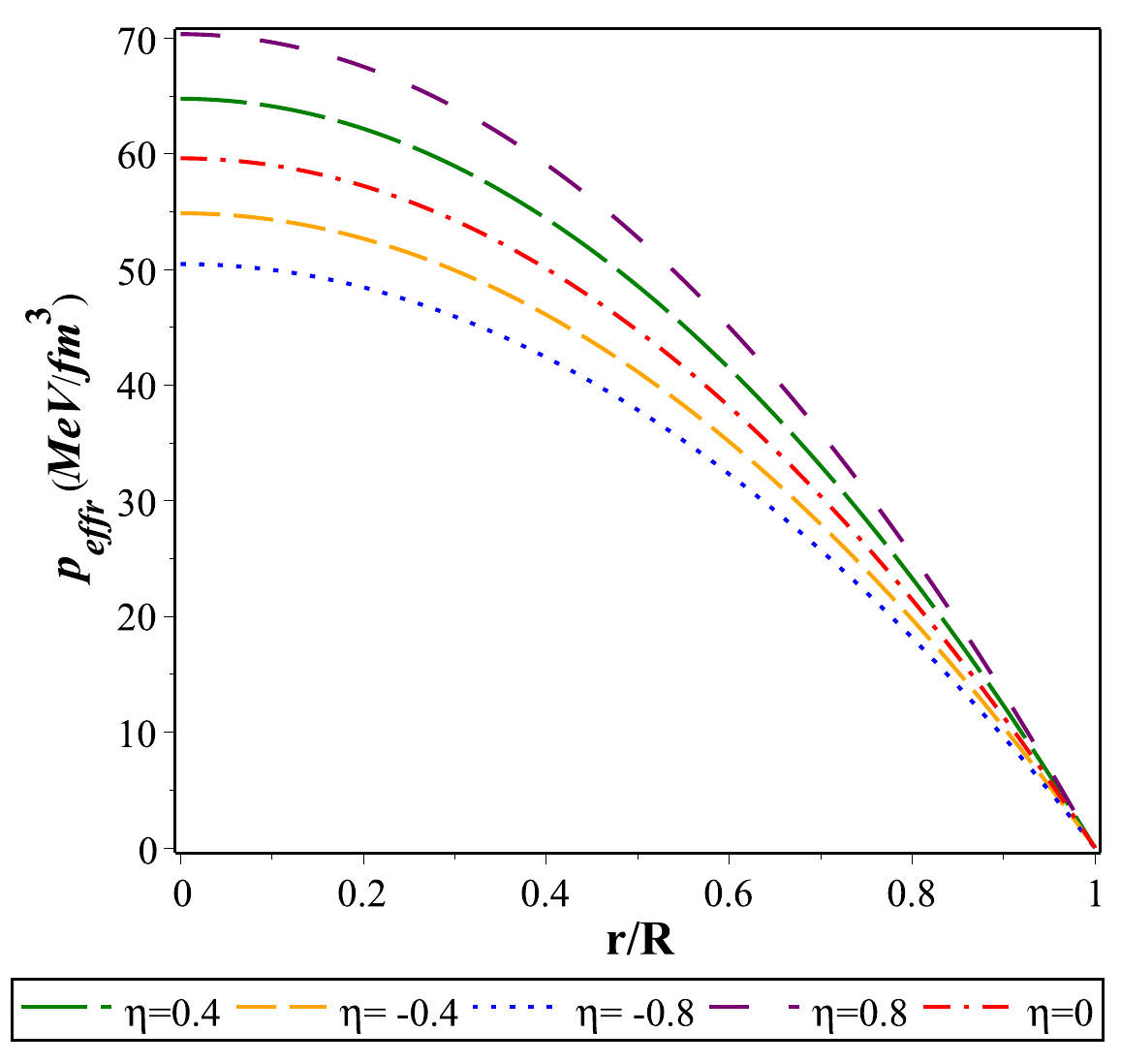}
	\caption{Variation of the radial pressure $(p_r)$ as a function of the fractional radial coordinate $r/R$, with bag constant (B$_g$) = 83~MeV/fm$^3$ and Finsler parameter ($\overline{Ric}$) = 1.2 for the $LMC~X-4$.} \label{fpr}
\end{figure}

Hence, the co-relation of the energy and pressure inside the strange stellar system can be interpreted as 
\begin{equation}
p_r=\frac{1}{3}(\rho-4B_g). \label{21}
\end{equation}

All the corrections required for the energy and pressure functions of SQM have been maintained by introducing {\it ad hoc} bag function.

The radial pressure must be on the surface of a stellar system, therefore from Eq. (\ref{21}), we can conclude
\[
\rho_0 = 4B_g,
\]
where $ \rho_0 $ is the surface density (i.e. at $r= R$).

Hence, the modified form is as follows
\begin{equation}
p_r=\frac{1}{3}(\rho-\rho_0). \label{22}
\end{equation}

Following Moraes et al.~\cite{Moraes}, we consider the tangential component of pressure inside the system is related to the matter density in the form
\begin{equation}
p_t = \rho c_1 +c_2.
\end{equation}

\begin{figure}[htp!]
	\includegraphics[scale=0.4]{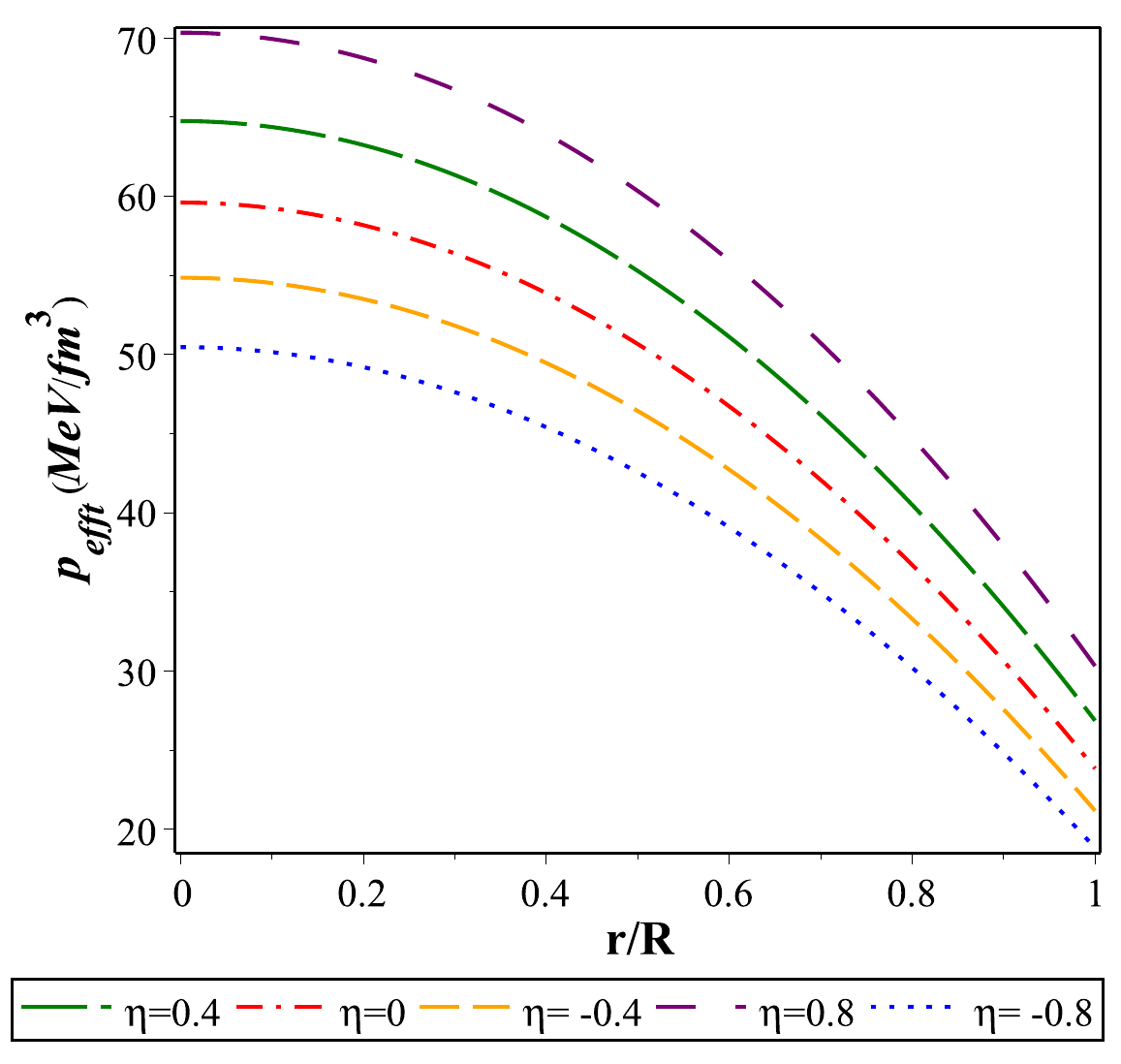}
	\caption{Variation of the tangential pressure $(p_t)$ as a function of the fractional radial coordinate $r/R$, with bag constant (B$_g$) = 83~MeV/fm$^3$ and Finsler parameter ($\overline{Ric}$) = 1.2 for the $LMC~X-4$.} \label{fpt}
\end{figure}

The variations of the physical quantities, like the radial and tangential pressure shown in Figs. \ref{fpr} and \ref{fpt} respectively in reference of the fractional radial coordinate for different coupling constant.

From the conservation equation of the stress-energy tensor, we obtain the hydrostatic equation of the strange stellar system in the following form 
\begin{equation}
-p_r'-\frac{\lambda'}{2}(\rho+p_r)+\frac{2}{r}(p_t-p_r) + \frac{\eta}{24 \pi_F} \big(3 \rho' - p'_r - 2 p'_t\big) = 0. \nonumber
\end{equation}

Now, following Eq. (\ref{16a}), the simplified form of the above equation can be written as
\begin{eqnarray}
p'_r = & &- \Biggl[ p_r \Big\{ (4 \pi_F r^3 + \frac{\eta r^3}{6}(\rho +p_r) + 2 r \overline{Ric} -3m) \Big\}  \nonumber \\ & &{\hspace{0.5cm}} -2p_t(r \overline{Ric} -m) - \frac{\eta r^3}{6}(3\rho -2p_t) +m \rho  \nonumber \\ & & - \frac{\eta r}{24 \pi_F}(3\rho' -2p'_t)(r \overline{Ric} -m) \Biggr] \Big/ (r \overline{Ric} -m) (r- \frac{\eta r}{24 \pi_F}). \nonumber \\ 
\end{eqnarray}

The expected result for the hydrostatic equilibrium condition in the Finslerian background for the strange stellar system can be obtained from $\eta = 0 $.

\section{Discussion and Conclusion} \label{dis}
In order to enhance the analysis of a feasible Finslerian generalization of the Einstein equations, we have developed an action-based Einstein field equation for the modified gravity (Eq. \ref{5}), evaluating the Finsler function of the Finsler spacetime. Our Finsler gravity theory incorporates the definition of the matter fields coupled with the Finsler spacetime by the principle which produces the necessary action from the Lagrangian norm on Lorentzian spacetime. We have obtained the Einstein field equations through variance about the basic function of geometry. It could suggest that, in the metric geometry limit, it becomes comparable to the Einstein field equations for coupling variable ($\eta$) = 0 \cite{Sourav}. To develop a physically stable system, we choose $\overline{Ric} \geq$ 1.

As a further formal development, we represent a model of the strange stellar system in the modified gravity background in the extension of the Finslerian structure. The modified density and pressure are developed from the coupled matter action, which depends on the coupling constant's behaviour.

\begin{figure}[htp!]
	\includegraphics[scale=0.4]{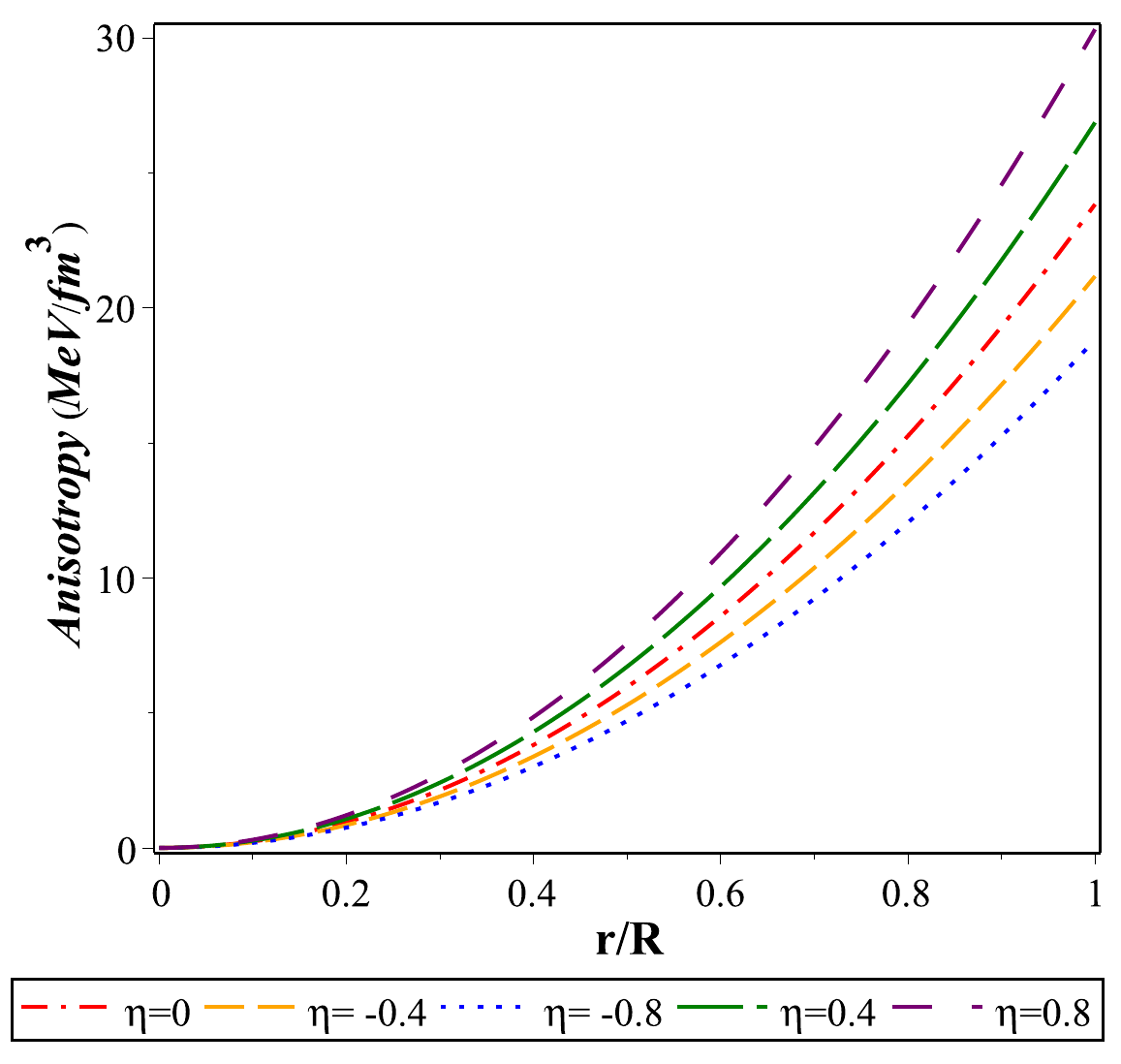}
	\caption{Variation of the anisotropic stress $(\Delta)$ as a function of the fractional radial coordinate $r/R$, with bag constant (B$_g$) = 83~MeV/fm$^3$ and Finsler parameter ($\overline{Ric}$) = 1.2 for the $LMC~X-4$.} \label{fani}
\end{figure}

The overall force operating is the anisotropic flow in addition to the differential pressure provided by the gravity operating on the shell by the substance (mass) within it. This determines the fluid element's hydrostatic equilibrium, which is at rest within the structure, as well as the overlying matter, which decreases with the radial coordinate. The difference in the stress of the tangential and radial component of pressures (anisotropic stress) is displayed in Fig. \ref{fani}. In particular, the anisotropic flow is shown to be well defined over the system and reaches the maximum at the surface of the stellar model. 

\begin{table*}[htp!]
	\centering \caption{\label{tab:table1} Numerical values of the physical parameters for different coupling constant {$\eta$} for the strange star $LMC~X-4$ of mass 1.29$M_\odot$ ($1~{{M}_{\odot}}=1.475~km$) with $\overline{Ric}$ =1.2 and B$ _g $ = 83 Mev/fm${^3}$.}
	\begin{tabular}{c  c  c  c  c  c  c}
		\hline \hline
		Value of $\eta$  &\hspace{0.55cm} $ \eta $ = -0.8 & \hspace{0.55cm} $ \eta $ = -0.4 & \hspace{.55cm} $ \eta $ = 0.0 &\hspace{0.55cm} $ \eta $ = 0.4 & \hspace{0.55cm} $ \eta $ = 0.8\\ \hline \hline\\  
		
		Predicted Radius ($km$) &\hspace{0.55cm} 9.946 &\hspace{0.55cm} 9.708 &\hspace{0.55cm}  9.475 &\hspace{0.55cm} 9.246 &\hspace{0.55cm}  9.021  \\
		
		$\rho_{effc}~({gm/cm}^3)$ &\hspace{0.55cm} $7.694\times 10^{14} $ &\hspace{0.55cm}  $8.366\times 10^{14} $ &\hspace{0.55cm} $9.106\times 10^{14} $ &\hspace{0.55cm} $9.923\times 10^{14} $ &\hspace{0.55cm}  $10.830 \times 10^{14} $ \\ 
		
		$\rho_{effo}~({gm/cm}^3)$ &\hspace{0.55cm} $5.237\times 10^{14} $ &\hspace{0.55cm}  $5.569\times 10^{14} $ &\hspace{0.55cm} $5.918\times 10^{14} $ &\hspace{0.55cm} $6.287\times 10^{14} $ &\hspace{0.55cm}  $6.674\times 10^{14} $ \\ 
		
		$P_{effc}~({dyne/cm}^2)$ &\hspace{0.55cm} $8.087\times 10^{34} $ &\hspace{0.55cm} $8.789\times 10^{34} $ &\hspace{0.55cm} $9.550\times 10^{34} $ &\hspace{0.55cm} $10.370\times 10^{34} $ &\hspace{0.55cm} $11.270\times 10^{34} $ \\ 
		
		$\frac{2M}{R}$  &\hspace{0.55cm} 0.38 &\hspace{0.55cm} 0.39 &\hspace{0.55cm} 0.40 &\hspace{0.55cm} 0.41 &\hspace{0.55cm} 0.42 \\ 
		
		Red Shift ($Z_s$) &\hspace{0.55cm} 0.27 &\hspace{0.55cm} 0.28 &\hspace{0.55cm} 0.29 &\hspace{0.55cm} 0.30 &\hspace{0.55cm} 0.31 \\ 
		\hline \hline
	\end{tabular}
\end{table*}

\begin{table*}
	\centering \caption{\label{tab:table2} Numerical values of the physical parameters for different {$\overline{Ric}$} for the strange star $LMC~X-4$ of mass 1.29$M_\odot$ ($1~{{M}_{\odot}}=1.475~km$) with coupling parameter $\eta$ =0.4 and B$ _g $ = 83 Mev/fm${^3}$.}
	\begin{tabular}{c  c  c  c  c  c  c}
		\hline \hline
		Value of $\overline{Ric}$  &\hspace{0.55cm} $ \overline{Ric} $ = 1 & \hspace{0.55cm} $ \overline{Ric} $ = 1.1 & \hspace{.55cm} $ \overline{Ric} $ = 1.2 \\ \hline \hline\\  
		
		Predicted Radius($km$) &\hspace{0.55cm} 9.456 &\hspace{0.55cm} 9.349 &\hspace{0.55cm}  9.246  \\
		
		$\rho_{effc}~({gm/cm}^3)$ &\hspace{0.55cm} $8.653\times 10^{14} $ &\hspace{0.55cm}  $9.284\times 10^{14} $ &\hspace{0.55cm} $9.923\times 10^{14} $ \\ 
		
		$\rho_{effo}~({gm/cm}^3)$ &\hspace{0.55cm} $6.731\times 10^{14} $ &\hspace{0.55cm}  $6.290\times 10^{14} $ &\hspace{0.55cm} $6.287\times 10^{14} $ \\ 
		
		$P_{effc}~({dyne/cm}^2)$ &\hspace{0.55cm} $6.731\times 10^{34} $ &\hspace{0.55cm} $8.541\times 10^{34} $ &\hspace{0.55cm} $10.370\times 10^{34} $ \\ 
		
		$\frac{2M}{R}$  &\hspace{0.55cm} 0.402 &\hspace{0.55cm} 0.407 &\hspace{0.55cm} 0.412 \\ 
		
		Red Shift($Z_s$) &\hspace{0.55cm} 0.293 &\hspace{0.55cm} 0.299 &\hspace{0.55cm} 0.304 \\ 
		
		\hline \hline
		
	\end{tabular}
\end{table*} 

\begin{figure}[htp!]
	\includegraphics[scale=0.45]{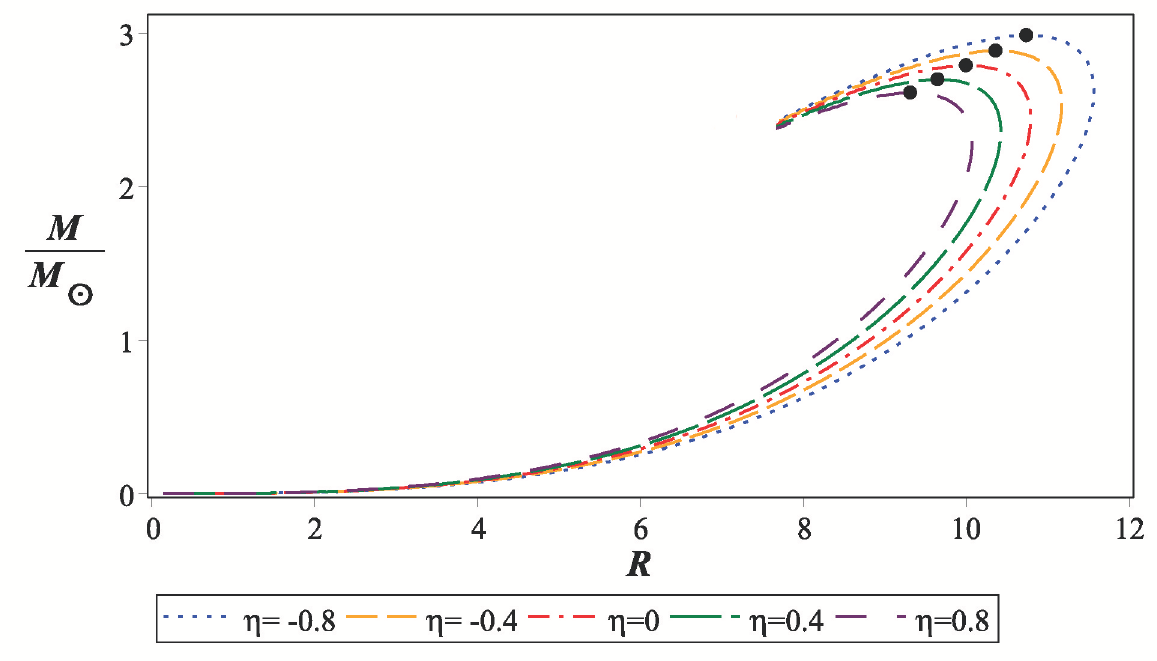}
	\caption{Variation of the mass of a strange star as a function of radius. Solid circles represent the maximum mass and radius of the respective curves. Here curves are drawn for $B_g$ = 83~MeV/fm$^3$ and Finsler parameter ($\overline{Ric}$) = 1.2.} \label{fmr}
\end{figure}

Regarding the variation of the total mass $ M $ in terms of normalized M$_\odot$ with respect to the radius for an opted value of $ \overline{Ric} $ = 1.2 and $ B_g=83 MeV/fm{^3} $ is presented in Fig. \ref{fmr}. The maximum mass point corresponds to the specific $\eta$, and the radius is marked by a solid circle. An increment of the $\eta$ increases the maximum mass and respective radius. We obtain the maximum mass for $\eta$ = 0 is 2.788 M$_\odot$, with a radius of 10.002 km. It is interesting to note that the mass decreases by 6.33\% and the corresponding radius reduces by 6.88\% for $\eta$ = 0.8. Further, we found that for $\eta$ = -0.8, the total mass rises by 7.06\% along with the radius increment by 7.43\%. By these, we can characterize that the higher coupling parameter compacted the stellar system. All variations for the mass-radius relation are suitable for the singularity condition.

\begin{figure}[htp!]
	\includegraphics[scale=0.4]{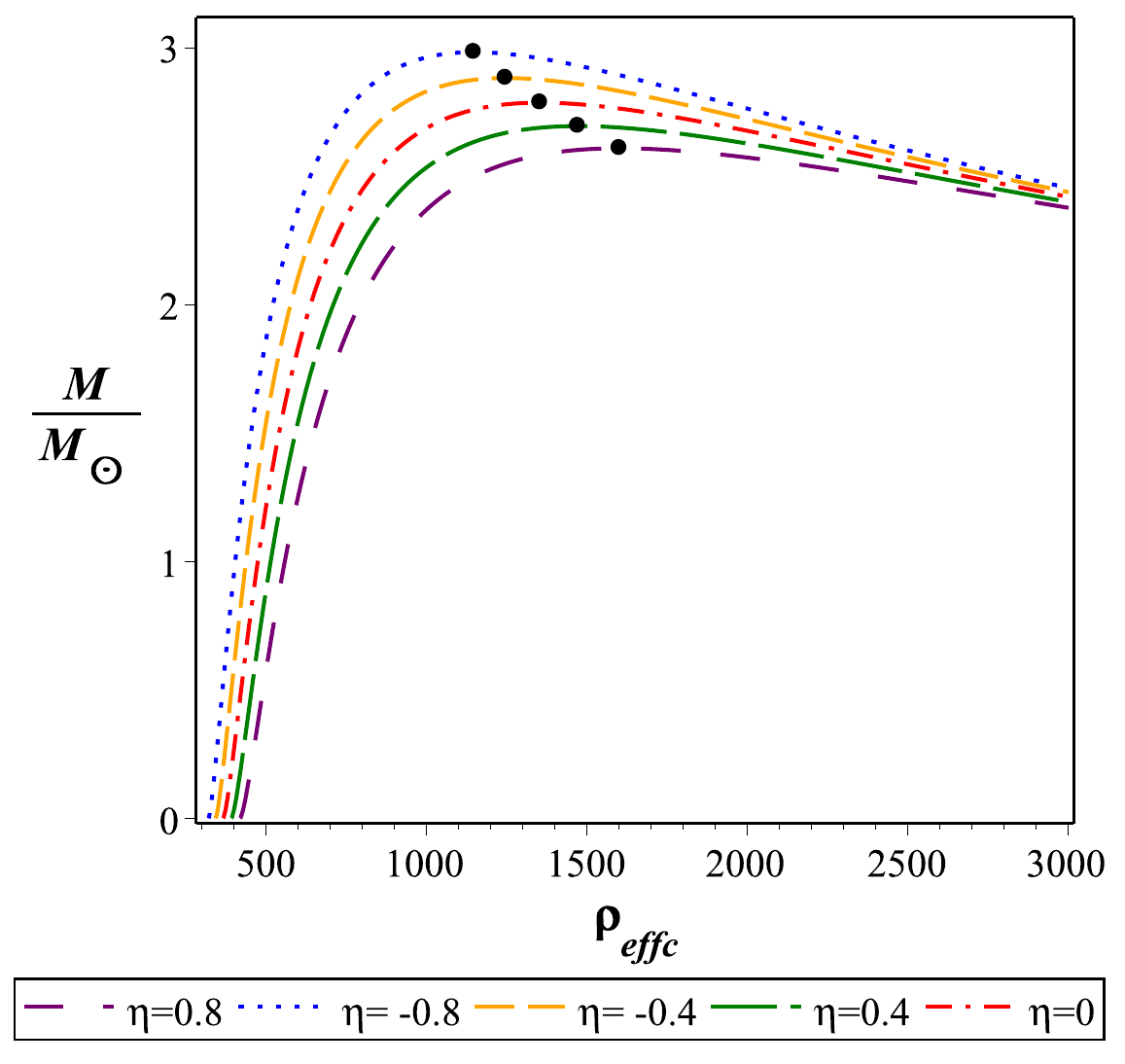}
	\caption{The variation of mass of a strange star as a function of the central density $({\rho}_c)$. Solid circles represent the maximum mass and radius of the respective curves. Curves are drawn for $B_g$ = 83~MeV/fm$^3$ and Finsler parameter ($\overline{Ric}$) = 1.2.}  \label{fmrho}
\end{figure}

The essential condition of a stellar system to be stable is $\frac{dM}{d \rho_c} > 0$. Variation of the stellar mass in M$_\odot$ with the central density $\rho_c$ is shown in Fig. \ref{fmrho}. The variation enunciates the central density attain the value 1355.445 $MeV/fm{^3} $ for the maximum mass 2.788 M$_\odot$ corresponds to $\overline{Ric}$ =1.2 and $\eta$ = 0, whereas the uttermost value of the central density are 1602.481 $MeV/fm{^3} $ and 1149.021 $MeV/fm{^3} $ for $\eta$ = 0.8 and -0.8 respectively. Complete circles over the curves show where the highest mass amounts are found with central density.  

The lower mass gap of 2.5--5 M$_\odot$, which lies between the heaviest known neutron star and the lightest defined black hole, has always been astonished scientists. In the recent observations by adv LIGO and Virgo, they have observed two events, GW200210 and GW190814 \cite{ligo1,ligo2,ligo3}, where the second companion masses are $2.83^{+0.47}_{-0.42} \textrm{~M}_\odot$ and $2.6^{+0.1}_{-0.1} \textrm{~M}_\odot$ respectively, lie in this mass gap. The mass gap might not actually exist but rather be a result of limits than something else. The secondary companion is typically thought to be a very light black hole due to the maximum mass restrictions of known nuclear EOS and the GR relativistic TOV equation. Depending on the coupling parameter and Finsler parameter, the modified gravity equation for a compact object combined with known EOS can result in a more massive stellar structure than GR. For further progress, we are working on the gravitational echoes and tidal deformity of the structure.

We have developed a concise study in the tabular format (Table \ref{tab:table1}) for the observed mass of $LMC-X4$ of different physical parameters. The study is developed for  B$ _g $ = 83 Mev/fm${^3}$ and $ \overline{Ric} =1.2$  under the chosen values of $\eta$ as -0.8, -0.4, 0.0 , 0.4 and 0.8. On the other hand, Table \ref{tab:table2} shows the diversity of the physical parameters with the variation of the Finsler parameter $\overline{Ric}$.

According to the present study, it is clear that with the increment of the coupling parameter as well as the Finsler parameter, the radii of the system become lesser, and the central density increases significantly, which depicts that the Finslerian background is a strong candidate to describe the compact system.

\appendix

\section*{Appendix: {\it Two Dimensional Finsler Space with constant flag}} \label{app}

The appearance of the Finsler spacetime with Lorentzian signature is the point of interest. On the Finsler manifold, every point of the Finsler structure $\mathcal{F}$ is not positive definite. $\mathcal{F}$=0 for the massless stipulation. Finsler space can be classified into two types: (i) Riemannian space, where $\mathcal{F}^2$ is quadratics in $ y $ and (ii) Randers space \cite{Randers}, where 
\begin{equation}
\mathcal{F}(x,y)\equiv  a (x,y) +b(x,y), \label{ran}
\end{equation}
with $a= \sqrt{\tilde{a}_{\mu \nu}(x)y^\mu y^\nu}$ where $\tilde{a}_{ij}$ is the Riemannian metric and $ b = \tilde{b}_\mu (x) y^\mu$ where, $\tilde{b}_\mu $ is 1 form.

Let us consider the isometric transformation of $ x $ under the infinitesimal coordinate transformation as follows
\begin{equation*}
\tilde{x}^{\mu} = x^{\mu} + \epsilon \tilde{V}^{\mu}.
\end{equation*}

The corresponding $ y $ transforms as
\begin{equation*}
\tilde{y}^{\mu} = y^{\mu} + \epsilon \frac{\partial \tilde{V}}{\partial x^{\nu}} y^{\nu},
\end{equation*}
with $|\epsilon|<<1$. 

The Finsler structure can be expanded by incorporating the transformations of $ x $ and $ y $. The expansion is considered up to the first order in $|\epsilon|$ as follows:
\begin{equation}
\tilde{\mathcal{F}}(\tilde{x},\tilde{y}) =  \tilde{\mathcal{F}}(x,y) + \epsilon \tilde{V}^{\mu} \frac{\partial \mathcal{F}}{\partial x^{\mu}}+ \epsilon y^{\nu}{\frac{\partial \tilde{V}^{\mu}}{\partial x^{\mu}}} {\frac{\partial \mathcal{F}}{\partial y^{\mu}}}.
\end{equation}

Using the expansion of the structure, the killing equation of the space can be expres as
\begin{equation}
K_V(\mathcal{F}) \equiv \tilde{V}^{\mu} \frac{\partial \mathcal{F}}{\partial x^{\mu}}+ y^{\nu}{\frac{\partial \tilde{V}^{\mu}}{\partial x^{\mu}}} {\frac{\partial \mathcal{F}}{\partial y^{\mu}}} = 0.
\end{equation}

By introducing the Randers length element defined in Eq. ($\ref{ran}$) and since the rational and irrational terms of the Killing equation are independent of each other, therefore 
\begin{eqnarray}
& & \tilde{V}_{\mu| \nu} + \tilde{V}_{\nu| \mu} = 0,\nonumber \\
& & \tilde{V}^{\mu} \frac{\partial \tilde{b}_\nu}{\partial x^{\mu}} + \tilde{b}_{\mu} \frac{\partial \tilde{V}^\mu}{\partial x^{\nu}} = 0.\\\nonumber
\end{eqnarray}

In Riemannian background, for a fixed radial coordinate $ r $, if the metric has the form $\mathcal{F}_{RS} = \sqrt{(y^\theta)^2 + \sin \eta (y^\phi)^2 }$ then the system can be consider as spherical symmetric with constant curvature. The ``Finslerian sphere" is the equivalent to the spherical symmetry of Riemannian space. The most of the celestial objects should hold the feature of spherical symmetry. The ``Finslerian sphere" also preserves the maximum possible symmetry. This is the topologically equivalent of a sphere from the mathematical definition. The flag curvature in Finsler geometry is generally the sectional curvature of the Riemannian frame. The constant Ricci scalar and the constant flag curvature are equivalent to each other.  A two-dimensional Finsler space has only one independent Killing vector \cite{Wang}.  Bao et al.~\cite{Bao2000} provide the two-dimensional Randers-Finsler space with constant positive flag curvature $\lambda$ = 1 as follows
\[\mathcal{F}_{FS} = \frac{\sqrt{(1-\epsilon^2 \sin ^2 \theta)y^\theta y^\theta + \sin ^2 \theta y^\phi y^\phi}}{1-\epsilon^2 \sin ^2 \theta} - \frac{\epsilon^2 \sin ^2 \theta ~y^\phi}{1-\epsilon^2 \sin ^2 \theta}. \]
with $0\leq \epsilon \leq 1$. For $\epsilon = 0$, the metric return to the Riemann sphere.

\section*{ACKNOWLEDGMENTS}
FR and SR are thankful to the Inter-University Centre for Astronomy and Astrophysics (IUCAA), Pune, India for providing Visiting Associateship under which a part of this work was carried out. FR is also grateful to the project DST-SERB (EMR/2016/000193), Government of India for providing financial support. SR acknowledges the supports and facilities under ICARD, Pune at CCASS, GLA University, Mathura.

\end{document}